\documentclass[a4,12pt]{article}

\usepackage{a4wide}
\usepackage{bm}% bold math
\usepackage{amsfonts}% bold math
\usepackage{amsmath}
\usepackage{graphics}

\makeatletter

\@addtoreset{equation}{section}
\makeatother

\def\d{\mathrm{d}}
\def\e{\mathrm{e}}
\def\im{\mathrm{i}}
\def\smalli{\mbox{\scriptsize \rm i}}
\def\sign{\varepsilon}

\title{\Large{\bf{Inverse Scattering Method for Square Matrix
Nonlinear Schr\"odinger Equation 
under Nonvanishing Boundary Conditions}}}

\author{Jun'ichi \textsc{Ieda}$^{1}$
{\footnote {\tt E-mail: ieda@imr.tohoku.ac.jp}},
\setcounter{footnote}{2}
Masaru \textsc{Uchiyama}$^{2}$ 
{\footnote {\tt E-mail: uchiyama@monet.phys.s.u-tokyo.ac.jp}}
\ \ and Miki \textsc{Wadati}$^{2}$\\
\\ \\
$^{1}${\it Institute for Materials Research, Tohoku University,} \\
{\it 2-1-1 Katahira, Aoba-ku, Sendai 980-8577, Japan}\\
\\
$^{2}${\it Department of Physics, Graduate School of Science,
University of Tokyo,}\\ {\it 7-3-1 Bunkyo-ku, Tokyo 113-0033, Japan}
}

\date{}

\begin{document}
\maketitle

\begin{abstract}
\baselineskip=6mm
Matrix generalization of the inverse scattering method is developed to solve the multicomponent nonlinear Schr\"odinger equation with nonvanishing boundary conditions. It is shown that the initial value problem can be solved exactly. 
The multi-soliton solution is obtained from the Gel'fand--Levitan--Marchenko equation.
\end{abstract}

\newpage

\baselineskip=7mm

\section{Introduction}
\label{sec:introduction}
The nonlinear Schr\"odinger (NLS) equation in the $(1+1)$ space-time dimension is one of the completely integrable systems, i.e., the soliton equations~\cite{refsoliton,FadTak}. This model equation has been extensively studied to describe nonlinear dynamics in a wide range of physics from fiber optics~\cite{fiberoptics,opticalsoliton1,opticalsoliton} to Bose--Einstein condensation of cold atoms~\cite{BECsoliton, Tsurumi,BECbsoliton_review}. The initial value problem can be solved exactly via the inverse scattering method (ISM)~\cite{Zak72,Zak73}. In particular, the reflection-free condition reduces the inverse problem to a set of algebraic equations making it possible to obtain the $N$-soliton solution in an explicit way.

One of major developments in the study of the NLS equation is multicomponent extensions preserving the integrability. Manakov~\cite{Manakov} studied a system of the coupled NLS (cNLS) equations on the basis of the ISM and obtained the soliton solutions. While the interaction of vector solitons for the multicomponent focusing NLS equation is elastic in the vector sense it was shown that during the two-soliton collision exchanges among components of each soliton may occur for particular choices of the parameter values \cite{Manakov,Tsuchida2}. In \cite{Tsuchida1}, the ISM for a matrix generalization of the NLS equation (in general, in a rectangular matrix form) was developed for solving the initial value problem. By assuming the reflection-free condition and \emph{vanishing boundary conditions} (see below) the $N$-soliton solution was obtained explicitly. It should be noted that, by appropriate identifications of the matrix-field elements, the matrix NLS equation reduces to the cNLS equations of the Manakov-type \cite{Tsuchida2,Tsuchida1} and remarkably to the spinor-type that is discovered recently~\cite{IMWlett,IMWfull} in connection with Bose--Einstein condensates with the spin degrees of freedom. Results given in \cite{Tsuchida1} for a general matrix-field, such as the $N$-soliton solution, conservation laws and Hamiltonian structure, are directly applicable to the reduced systems. Thus, a further analysis of the matrix NLS equation is desired to deal with multicomponent nonlinear dynamics under different circumstances.

In this paper, we study a square matrix NLS equation under the \emph{nonvanishing boundary conditions} by means of the ISM. The multicomponent system with such boundary conditions is regarded as an extension of a basic single-component NLS equation with the self-defocusing nonlinearity studied by Zakharov and Shabat~\cite{Zak73} and also that with the self-focusing nonlinearity investigated by Kawata and Inoue~\cite{KawIno78}. As compared to the case with the vanishing boundary conditions~\cite{Tsuchida1}, the conservation laws and Hamiltonian structure are the same, while the Lax pair, the direct and inverse problems, and the $N$-soliton solution should be reformulated reflecting the boundary values of the matrix-field.

This paper is organized as follows. In Sec.~\ref{sec:Formulation}, nonvanishing boundary conditions for the matrix NLS equation are introduced. The Lax pair is provided to formulate the auxiliary linear system. Then the conservation laws are constructed systematically. In Sec.~\ref{sec:ISM}, the direct and inverse problems are solved along the ISM procedure and the multi-soliton solution is presented. Section \ref{sec:conclusions} is devoted to the concluding remarks.

\section{Formulation}
\label{sec:Formulation}
The matrix NLS equation is expressed as
\begin{align}
\label{eq:genNLS}
\im Q_t+Q_{xx}-2 \sign QQ^\dagger Q=O\,\quad(\sign =\pm1),
\end{align}
where $Q(x,t)$ and $O$ are an $l\times l$ matrix valued function and the zero matrix, 
respectively, $Q^\dagger$ is the Hermitian conjugate of $Q$, 
and the subscripts $t$, $x$ denote the partial derivatives. 

The case $\sign=-1$ ($\sign=+1$) of (\ref{eq:genNLS}) is often referred to as 
the self-focusing (-defocusing) one. 
In \cite{Tsuchida1}, it was shown that through the ISM the system has 
an infinite number of conservation laws. The initial value problem was solved and 
the $N$-soliton solution was obtained under the constraint $\sign=-1$ and 
the vanishing boundary condition:
\begin{equation}
Q(x,t)\to O \,\,\mbox{ as }\,\, x\to\pm\infty.
\end{equation}
Under these conditions each soliton forms the so-called bright soliton with 
$l^2$ components.
Setting the form 
\begin{align}
Q=\left(\begin{array}{ccc}
Q_{11} & \cdots & Q_{1l}\\
\vdots &  & \vdots\\
Q_{m1} & \cdots & Q_{ml}\\
0 & \cdots & 0\\
\vdots &  & \vdots\\
0 & \cdots & 0
\end{array}
\right),
\label{eq:reduction}
\end{align}
with $m<l$, one can achieve a rectangular matrix reduction that is compatible with the vanishing boundary condition.
In particular, the $m=1$ case corresponds to the $l$-component Manakov model~\cite{Tsuchida2}. 
Other types of reduction can be obtained in a nontrivial way 
by putting some components equal without breaking consistency of equations~\cite{IMWlett,IMWfull}.

On the other hand, for another integrable constraint $\sign=+1$, leading to
\begin{equation}
\label{eq:mtxNLS+}
\im Q_t+Q_{xx}-2 QQ^\dagger Q=O,
\end{equation}
the boundary condition should be altered appropriately, which has not been 
investigated so far. 
Equation (\ref{eq:mtxNLS+}) is a matrix generalization of the NLS equation for 
a scalar field $q(x,t)$ with a self-defocusing nonlinearity, 
\begin{equation}
\label{eq:NLS+}
\im q_t+q_{xx}-2 |q|^2q=0,
\end{equation}
which possesses dark soliton solutions.
For the self-defocusing NLS equation (\ref{eq:NLS+}), the boundary condition at 
$x\to\pm\infty$ is assumed to be the nonvanishing one, e.g., a constant, 
$|q(x)|\to \lambda_0$, rather than the vanishing one, $|q(x)|\to 0$.
The ISM procedure was applied to the system (\ref{eq:NLS+}) in \cite{Zak73}. 
The analysis of the NLS equation under the nonvanishing boundary conditions was 
extended to the self-focusing case in \cite{KawIno78}.

From now on, we concentrate on the analysis of a full-rank $l\times l$ square matrix NLS equation. 
We do not include reductions (\ref{eq:reduction}) in this work, i.e., the vector
(Manakov) NLS equation falls outside our considerations.
Systematic study of such reductions based on the symmetry argument is an open issue.
In this section, we introduce a square matrix type of nonvanishing boundary 
conditions for eq.~(\ref{eq:genNLS}) and formulate the Lax equation for the ISM.

\subsection{Nonvanishing boundary condition}
\label{ssec:NBC}
We assume that the $l\times l$ matrix valued function $Q(x,t)$ satisfies the following 
nonvanishing conditions,
\begin{align}
\label{eq:constBC}
&Q(x,t)\to Q_\pm \,\,\mbox{ as }\,\, x\to\pm\infty,\\
\label{eq:constBCa}
&Q_\pm^\dagger Q_\pm= Q_\pm Q_\pm^\dagger=\lambda_0^2I,
\end{align}
where $\lambda_0$ is a positive real constant and $I$ denotes the $l\times l$ unit matrix.
Noting the freedom of unitary transformations:
\begin{align}
\label{eq:unitarytransf}
Q'=\mathcal{U}Q\mathcal{V},
\end{align}
with $\mathcal{U}$, $\mathcal{V}$ unitary matrices, we see that 
if $Q$ is a solution of eq.~(\ref{eq:genNLS}), then $Q'$ is also a solution. 
By this freedom, without loss of generality, we can fix one side of 
the boundary condition as
\begin{align}
\label{eq:constBC+}
Q_+=\lambda_0\e^{\im(kx-\omega t)}I.
\end{align}

To avoid a complexity, separate the carrier wave part,
\begin{align}
\label{eq:changeQ}
Q(x,t)= \hat Q(x,t)\e^{\smalli(kx-\omega t)},
\end{align}
where the dispersion relation is determined as
\begin{equation}
\label{eq:dispersion}
\omega=k^2+2\sign\lambda_0^2.
\end{equation}
Then the original nonlinear evolution equation (\ref{eq:genNLS}) is equivalent to
\begin{equation}
\label{eq:mtxNLS+tilde}
\im Q_t+Q_{xx}+2\im kQ_{x}+2\sign(\lambda_0^2I-QQ^\dagger)Q=O.
\end{equation}
Here and hereafter except the final expression (\ref{eq:finalNsoliton}), 
we drop the hat of $\hat Q$ for a notational simplicity. 
Accordingly, eq.~(\ref{eq:constBC+}) is rewritten as 
\begin{align}
Q_+=\lambda_0I.
\label{eq:constBC2}
\end{align}
In what follows, we focus on eq.~(\ref{eq:mtxNLS+tilde}) with the boundary conditions 
(\ref{eq:constBC}), (\ref{eq:constBCa}), and (\ref{eq:constBC2}).

\subsection{Lax pair}
We introduce the Lax matrices in the following forms,
\begin{align}
\label{eq:LaxU}
U&=\im\lambda \left[\begin{array}{cc}-I&O\\O&I\end{array}\right]
+\left[\begin{array}{cc}O&Q\\ \sign Q^\dagger&O\end{array}\right],\\
\label{eq:LaxV}
V&=\im\lambda^2\left[\begin{array}{cc}-2I&O\\O&2I\end{array}\right]
+\lambda\left[\begin{array}{cc}2\im k I&2Q\\ 2\sign Q^\dagger&-2\im k I\end{array}\right]
+\im\left[\begin{array}{cc}\sign(\lambda_0^2I-QQ^\dagger) & Q_x+2\im kQ\\
\sign(-Q^\dagger_x+2\im kQ^\dagger) & -\sign(\lambda_0^2I-Q^\dagger Q)\end{array}\right],
\end{align}
where $\lambda$ is the spectral parameter that is independent of time, $\lambda_t=0$. 
The potential matrix $Q$ satisfies the nonvanishing boundary conditions 
(\ref{eq:constBC}), (\ref{eq:constBCa}) with eq.~(\ref{eq:constBC2}).
In the ISM, we associate a set of linear problems: 
\begin{align}
\Psi_x=U\Psi, \qquad \Psi_t=V\Psi,\label{eq:linprob}
\end{align}
where $\Psi$ is a $2l\times l$ matrix function. 
We also use the following representations, 
\begin{align}
\Psi=\left[\begin{array}{c}\Psi_1\\ \Psi_2\end{array}\right], \quad
V=\left[\begin{array}{cc}V_{11}&V_{12}\\ V_{21}&V_{22}\end{array}\right],
\end{align}
where all the entries are $l\times l$ matrices.
Then, the Lax equation is obtained from the compatible condition of 
eqs. (\ref{eq:linprob}),
\begin{equation}
U_t-V_x+UV-VU=O,
\label{eq:laxeq}
\end{equation}
which is equivalent to the matrix NLS equation (\ref{eq:mtxNLS+tilde}).

\subsection{Conservation laws}
In this subsection, we recapture a systematic method to construct local conservation 
laws for the matrix NLS equation~\cite{Tsuchida1}. 
The method was originally developed for the scalar field case \cite{WadatiNLSE}. 

Introduce the quantity, 
\begin{align}
\Gamma=\Psi_2\Psi_1^{-1}.
\end{align}
From eqs.~(\ref{eq:LaxU})--(\ref{eq:laxeq}), one can prove
\begin{align}
\label{eq:continuityeq}
\frac{\partial}{\partial t}{\rm tr}(Q\Gamma) 
=\frac{\partial}{\partial x} {\rm tr}(V_{12}\Gamma+V_{11}),\\
2\im\lambda Q\Gamma=-\sign QQ^\dagger+Q\Gamma_x+(Q\Gamma)^2.
\label{eq:QGamma}
\end{align}
Note that eq.~(\ref{eq:continuityeq}) has a form of conservation law. 
Expand $Q\Gamma$ in $\lambda$ as
\begin{align}
Q\Gamma=\sum_{j=1}^\infty \frac{\sign}{(2\im\lambda)^j} F_j.
\end{align}
Then the trace of each coefficient $F_j$ is a conserved density and 
eq.~(\ref{eq:continuityeq}) represents an infinite number of continuity equations. 
From eq.~(\ref{eq:QGamma}), we recursively obtain 
\begin{align}
&F_1=-QQ^\dagger,\\
&{\rm tr}F_2={\rm tr}(-QQ_x^\dagger),\\
&{\rm tr}F_3={\rm tr}(-QQ_{xx}^\dagger+\sign QQ^\dagger QQ^\dagger).
\end{align}
By direct calculation, all elements of $F_1$ are shown to be conserved densities. 
%Under the nonvanishing boundary conditions, the conserved quantities may be defined 
%in a hole picture, i.e., by subtracting these densities from those at the boundary and 
%integrating them over whole space.

\section{Inverse Scattering Method}
\label{sec:ISM}
In this section, we carry out the ISM procedure for eq.~(\ref{eq:mtxNLS+tilde}) 
based on the Lax pair (\ref{eq:LaxU}) and (\ref{eq:LaxV}).
\subsection{Direct problem}
\label{sec:direct}
We consider the eigenvalue problem,
\begin{equation}
\Psi_x=U\Psi,\qquad
U=\left[\begin{array}{cc}-\im\lambda I&Q\\ \sign Q^\dagger&\im\lambda I
\end{array}\right],
\label{eq:mtxeigenU}
\end{equation}
and define the Jost functions and scattering data for them. 
Here $Q$ plays a role of potentials in the eigenvalue problem.
In Sec. \ref{sec:direct} and \ref{sec:inverse}, the analysis will be made with 
fixed time~$t$. Under the nonvanishing boundary conditions (\ref{eq:constBC}),
$U$ has the asymptotic forms:
\begin{equation}
\label{eq:Uasympt}
U(\lambda)\to U^{\pm}(\lambda)=
\left[\begin{array}{cc}-\im\lambda I&
Q_\pm \\
\sign Q_\pm^\dagger&\im\lambda I
\end{array}\right]\quad \mbox{as}\quad x\to \pm\infty.
\end{equation}
The characteristic roots of $U^\pm(\lambda)$ are twofold,
\begin{equation}
\label{eq:root}
\im\zeta,\quad-\im\zeta,
\end{equation}
where $\zeta\equiv(\lambda^2-\sign\lambda_0^2)^{1/2}$.
We introduce $2l\times2l$ matrix functions $T$ and $T^\pm$ by
\begin{align}
T(\lambda,\zeta;x)&=
\left[\begin{array}{cc}-\im\widetilde{Q}(x)&(\lambda-\zeta)I\\
(\lambda-\zeta)I&\im\sign\widetilde{Q}^\dagger(x)
\end{array}\right],\label{eq:defofT}\\
T^\pm(\lambda,\zeta)&=\lim_{x\to\pm\infty} T(\lambda,\zeta;x).
\label{eq:T+-}
\end{align}
Here $\widetilde{Q}(x)$ is an $l\times l$ smooth matrix function
that satisfies the same boundary condition as eq. (\ref{eq:constBC}), 
\begin{equation}
\widetilde{Q}(x)\to Q_\pm
\qquad\mbox{ as }\,x\to\pm\infty ,
\end{equation}
and for all $x$, 
\begin{align}
\widetilde{Q}(x)\widetilde{Q}^\dagger(x)=\widetilde{Q}^\dagger(x)\widetilde{Q}(x)
=\lambda_0^2 I.
\end{align}
Using $T^{\pm}(\lambda,\zeta)$, we can diagonalize $U^{\pm}(\lambda)$
as follows,
\begin{equation}
\label{eq:diagU}
U^{\pm}(\lambda)=-\im\zeta T^{\pm}(\lambda,\zeta)(\sigma^z\otimes I)
[T^{\pm}(\lambda,\zeta)]^{-1},
\end{equation}
where $\sigma^i$ ($i=x,y,z$) is the Pauli matrix and 
$\otimes$ denotes the direct product,
\begin{equation}
\sigma^z\otimes I=\left[\begin{array}{cc}I&O\\O&-I\end{array}\right].
\end{equation}
By use of eqs.~(\ref{eq:diagU}), we define matrix Jost functions
$\psi^-_1$, $\psi^-_2$, $\psi^+_1$, and $\psi^+_2$
as solutions of eq.~(\ref{eq:mtxeigenU}), whose asymptotic forms
are, respectively, given by
\begin{subequations}
\label{eq:jost+}
\begin{align}
\label{eq:jost1+}
\psi_1^-\sim&T^{-}(\lambda,\zeta)\left[\begin{array}{c}
I \\
O
\end{array}\right]\e^{-\smalli\zeta x}\quad
\mbox{as}\quad x\to -\infty, \\
\label{eq:jost2+}
\psi_2^-\sim&T^{-}(\lambda,\zeta)\left[\begin{array}{c}
O \\
I
\end{array}\right]\e^{\smalli\zeta x}\quad
\mbox{as}\quad x\to -\infty, \\
\label{eq:jost3+}
\psi_1^+\sim&T^{+}(\lambda,\zeta)\left[\begin{array}{c}
I \\
O
\end{array}\right]\e^{-\smalli\zeta x}\quad
\mbox{as}\quad x\to +\infty, \\
\label{eq:jost4+}
\psi_2^+\sim&T^{+}(\lambda,\zeta)\left[\begin{array}{c}
O \\
I
\end{array}\right]\e^{\smalli\zeta x}\quad
\mbox{as}\quad x\to +\infty.
\end{align}
\end{subequations}
We note that $\{\psi_1^+,\psi_2^+\}$ as well as $\{\psi_1^-,\psi_2^-\}$ 
constitute fundamental systems of solution. 
This is easily proved by using the usual Wronskian defined by the determinant. 
In fact, one can show
\begin{equation}
\frac{\d}{\d x}\det[\Phi_1, \Phi_2]=0,
\end{equation}
for any two $2l\times l$ matrix solutions $\Phi_1$, $\Phi_2$ of eq.~(\ref{eq:mtxeigenU}). 
Checking the value at $x\to\pm\infty$, we have 
\begin{equation}
\det[\psi_1^\pm, \psi_2^\pm]=\left(2\zeta(\lambda-\zeta)\right)^l.%\neq 0,
\end{equation} 
This indicates the linear independence of $\psi_1^\pm$ and $\psi_2^\pm$ 
except the branch points of $\zeta$, i.e., $\lambda=\pm\sqrt{\varepsilon}\lambda_0$ at which the solutions degenerate. 

If we use a notation $\bm{\psi}^{\pm}\equiv[\psi_1^\pm,\psi_2^\pm]$, 
relations (\ref{eq:jost+}) can be rewritten compactly in the following form,
\begin{equation}
\label{eq:jostasympt+}
\bm{\psi}^{\pm}(\lambda,\zeta;x)\to T^{\pm}(\lambda,\zeta)J(\zeta x)
\quad \mbox{as}\quad x\to \pm\infty,
\end{equation}
where
\begin{equation}
J(\zeta x)\equiv \left[\begin{array}{cc}
\e^{-\smalli\zeta x}I&O\\O&\e^{\smalli\zeta x}I\end{array}\right].
\end{equation}
Then the scattering matrix $S(\lambda,\zeta)$ is defined by
\begin{align}
\label{eq:defscatmtx}
\bm{\psi}^{-}(\lambda,\zeta;x)&=\bm{\psi}^{+}(\lambda,\zeta;x)
S(\lambda,\zeta),\\
\label{eq:scatmtx}
S(\lambda,\zeta)&=\left[\begin{array}{cc}
A(\lambda,\zeta)&\bar{B}(\lambda,\zeta) \\
B(\lambda,\zeta)&\bar{A}(\lambda,\zeta)
\end{array}\right],
\end{align}
where all the entries $A$, $\bar{A}$, $B$, and $\bar{B}$ 
represented by $l\times l$ matrices constitute scattering data.

It is useful to consider another slightly modified eigenvalue problem~\cite{KI77}. 
Under a transformation,
\begin{equation}
\Phi\equiv T^{-1}\Psi,
\end{equation}
the eigenvalue problem (\ref{eq:mtxeigenU}) becomes
\begin{align}
\label{eq:mtxeigenUT}
\Phi_x= \widetilde{U}\Phi, \qquad
%T^{-1}(UT-T_x)\Phi.
\widetilde{U}\equiv T^{-1}(UT-T_x)
=-\im\zeta\sigma^z\otimes I+W,
\end{align}
where $W=(W_{ab})$, $a, b=1,2$ with $l\times l$ matrices, 
\begin{subequations}
\begin{align}
W_{11}&=\frac{\im\sign}{2\zeta}(Q^\dagger\widetilde{Q}+
\widetilde{Q}^\dagger Q-2\lambda_0^2 I)
-\frac{\sign}{2\zeta(\lambda-\zeta)}\widetilde{Q}^\dagger\widetilde{Q}_x,
\\
W_{12}&=-\frac{\sign}{2\lambda_0^2} \widetilde{Q}^\dagger\left[
\frac{\lambda}{\zeta}(\widetilde{Q}Q^\dagger +Q\widetilde{Q}^\dagger
-2\lambda_0^2 I)
+(Q\widetilde{Q}^\dagger-\widetilde{Q}Q^\dagger)\right]
+\frac{\im\sign}{2\zeta}\widetilde{Q}^\dagger_x,
\\
W_{21}&=-\frac{1}{2\lambda_0^2} \widetilde{Q}\left[
\frac{\lambda}{\zeta}(\widetilde{Q}^\dagger Q+Q^\dagger\widetilde{Q}
-2\lambda_0^2 I)
+(Q^\dagger\widetilde{Q}-\widetilde{Q}^\dagger Q)\right]
-\frac{\im}{2\zeta}\widetilde{Q}_x,
\\
W_{22}&=-\frac{\im\sign}{2\zeta}(\widetilde{Q}Q^\dagger+Q\widetilde{Q}^\dagger
-2\lambda_0^2 I)
-\frac{\sign}{2\zeta(\lambda-\zeta)}\widetilde{Q}\widetilde{Q}^\dagger_x.
\end{align}
\end{subequations}
As solutions of eq.~(\ref{eq:mtxeigenUT}), 
we introduce new Jost matrices $\bm{\phi}^{\pm}\equiv[\phi_1^\pm,\phi_2^\pm]=
T^{-1}\bm{\psi}^\pm$ 
with simpler asymptotic forms,
\begin{equation}
\label{eq:phiasymptotics}
\bm{\phi}^{\pm}(\lambda,\zeta;x)\to J(\zeta x)
\quad \mbox{as}\quad x\to \pm\infty.
\end{equation}
These matrix Jost functions are connected with each other
through the same scattering matrix (\ref{eq:scatmtx}) as
\begin{equation}
\label{eq:defscatmtx2}
\bm{\phi}^{-}(\lambda,\zeta;x)=\bm{\phi}^{+}(\lambda,\zeta;x)
S(\lambda,\zeta).
\end{equation}

We now examine the properties of the scattering data. 
For this purpose, define a matrix function for two $2l\times l$ matrix functions 
$\Phi_i(\lambda,\zeta;x)$ ($i=1,2$) as
\begin{equation}
\label{eq:wronskian}
M[\Phi_1,\Phi_2]\equiv\Phi_1^\dagger (\lambda^*,\zeta^*;x) \left[
\begin{array}{cc}
I&O\\
O&-\sign I
\end{array}
\right]\Phi_2 (\lambda,\zeta;x).
\end{equation}
If  $\Phi_1(\lambda,\zeta;x)$, $\Phi_2(\lambda,\zeta;x)$ are solutions of 
eq.~(\ref{eq:mtxeigenU}), one can show that 
\begin{equation}
\frac{\d}{\d x}M[\Phi_1,\Phi_2]=O.
\end{equation}
From the asymptotic forms (\ref{eq:jost+}), we obtain the following relations,
\begin{subequations}
\label{eq:wronskis+}
\begin{align}
&M[\psi_1^\pm,\psi_1^\pm]
=-\sign M[\psi_2^\pm,\psi_2^\pm]=2\sign\zeta(\lambda-\zeta)I,\\
&M[\psi_1^\pm,\psi_2^\pm]=O,\\
\label{eq:Azeta+}
&M[\psi_1^+,\psi_1^-]=2\sign\zeta(\lambda-\zeta) A(\lambda,\zeta),\\
%&A(\lambda,\zeta)=W[\phi_1^+,\phi_1^-]/2\sign\zeta(\lambda-\zeta),\\
\label{eq:Abarzeta+}
&M[\psi_2^+,\psi_2^-]=-2\zeta(\lambda-\zeta) \bar{A}(\lambda,\zeta),\\
%&\bar{A}(\lambda,\zeta)=-\sign W[\phi_2^+,\phi_2^-],\\
&M[\psi_2^+,\psi_1^-]=-2\zeta(\lambda-\zeta) B(\lambda,\zeta),\\
%&B(\lambda,\zeta)=-\sign W[\phi_2^+,\phi_1^-],\\
&M[\psi_1^+,\psi_2^-]=2\sign\zeta(\lambda-\zeta) \bar{B}(\lambda,\zeta).
%&\bar{B}(\lambda,\zeta)=W[\phi_1^+,\phi_2^-].
\end{align}
\end{subequations}
These relations are rewritten into 
\begin{align}
\label{eq:ABinv+}
\left[\begin{array}{cc}
A^\dagger(\lambda^*,\zeta^*)&-\sign B^\dagger(\lambda^*,\zeta^*)\\
-\sign \bar{B}^\dagger(\lambda^*,\zeta^*)&\bar{A}^\dagger(\lambda^*,\zeta^*)
\end{array}\right]\left[\begin{array}{cc}
A(\lambda,\zeta)&\bar{B}(\lambda,\zeta)\\
B(\lambda,\zeta)&\bar{A}(\lambda,\zeta)\end{array}\right]=
\left[\begin{array}{cc}I&O\\O&I\end{array}\right],
\end{align}
which can be shown, e.g., as
\begin{align}
M[\psi_1^-,\psi_1^-]
=&A^\dagger(\lambda^*,\zeta^*)M[\psi_1^+,\psi_1^+]A(\lambda,\zeta)+
B^\dagger(\lambda^*,\zeta^*)M[\psi_2^+,\psi_2^+]B(\lambda,\zeta)\nonumber\\
=&2\sign\zeta(\lambda-\zeta)(A^\dagger(\lambda^*,\zeta^*)A(\lambda,\zeta)
-\sign B^\dagger(\lambda^*,\zeta^*)B(\lambda,\zeta))\nonumber\\
=&2\sign\zeta(\lambda-\zeta)I.
\end{align}
The relation (\ref{eq:ABinv+}) leads to the inversion of eq.~(\ref{eq:defscatmtx}),
\begin{align}
\label{eq:defscatmtxinv}
\bm{\psi}^{+}(\lambda,\zeta;x)&=\bm{\psi}^{-}(\lambda,\zeta;x)
[S(\lambda,\zeta)]^{-1},\\
[S(\lambda,\zeta)]^{-1}&=\left[\begin{array}{cc}
A^\dagger(\lambda^*,\zeta^*)&-\sign B^\dagger(\lambda^*,\zeta^*)\\
-\sign \bar{B}^\dagger(\lambda^*,\zeta^*)&\bar{A}^\dagger(\lambda^*,\zeta^*)
\end{array}\right].
\end{align}
In this system, we have involution relations for the Jost functions 
\begin{align}
\bm{\psi}^\pm(\lambda,\zeta;x)=
\bm{\psi}^\pm(\lambda,-\zeta;x) \mathcal{P}^\pm(\lambda,\zeta),
\end{align}
where 
\begin{align}
\mathcal P^\pm(\lambda,\zeta)&=J(\zeta x)\left[T^\pm(\lambda,-\zeta)\right]^{-1}
T^\pm(\lambda,\zeta)J(\zeta x)\nonumber\\
&=\frac{1}{\lambda+\zeta}\left[
\begin{array}{cc}
O&\im \sign Q_\pm^\dagger\\
-\im Q_\pm &O
\end{array}
\right].
\end{align}
From the involution, we have another set of relations for the scattering data: 
\begin{align}
\label{eq:parity1}
&S(\lambda,\zeta)=\left[\mathcal P^+(\lambda,\zeta)\right]^{-1}S(\lambda,-\zeta)
\mathcal P^-(\lambda,\zeta),\\
\label{eq:parity2}
&\bar{A}(\lambda,\zeta)=\frac{1}{\lambda_0^2}Q_+ A(\lambda,-\zeta)Q_-^\dagger,\\
&\bar{B}(\lambda,\zeta)=-\frac{\sign}{\lambda_0^2}Q_+^\dagger B(\lambda,-\zeta)Q_-^\dagger.
\end{align}
By use of eqs.~(\ref{eq:defscatmtx}) and (\ref{eq:defscatmtxinv}), we can prove 
\begin{align}
\det A(\lambda,\zeta)&=\det[\psi_1^-\ \psi_2^+]=\det \bar{A}^\dagger(\lambda^*,\zeta^*)
\nonumber\\
&=\left(\det \bar{A}(\lambda^*,\zeta^*)\right)^*.
\label{eq:reldetA1}
\end{align}
Additionally, taking the determinant of both sides of eq.~(\ref{eq:parity2}), we have 
\begin{align}
\det \bar{A}(\lambda,\zeta)=
\det A(\lambda,-\zeta)\det\left[\lambda_0^{-2}Q_+Q_-^\dagger\right].
\label{eq:reldetA2}
\end{align}
Combining eqs.~(\ref{eq:reldetA1}) and (\ref{eq:reldetA2}), we find that 
$\det A(\lambda,\zeta)\propto\left(\det A(\lambda^*,-\zeta^*)\right)^*$. 
As a consequence, if $(\lambda_j,\zeta_j)$ is the zero of $\det A$, 
$(\lambda_j^*,-\zeta_j^*)$ is also the zero of $\det A$, and 
the pairs $(\lambda_j,-\zeta_j)$ and $(\lambda_j^*,\zeta_j^*)$ 
are the zeros of $\det \bar{A}$. For $\sign=+1$, furthermore, 
the self-adjointness of the eigenvalue problem (\ref{eq:mtxeigenU}) leads to 
$\lambda_j=\lambda_j^*\in\mathbb{R}$ and $\zeta_j=-\zeta_j^*\in\im\mathbb{R}$. 

Finally, we make clear the analytical properties of
the Jost functions (\ref{eq:jost+}) in regard to complex $\lambda$.
To this end, we prepare a two-sheet Riemann surface for $\lambda$ 
where $\zeta\equiv(\lambda^2-\sign\lambda_0^2)^{1/2}$ is single-valued. 
For $\sign=+1$, cuts are made in $(-\infty,-\lambda_0]$ and $[\lambda_0,\infty)$
(see Fig.\ref{fig:contour+}). Each sheet is characterized such that 
$\textrm{Im}\ \zeta>0$ ($\textrm{Im}\ \zeta<0$) on the upper (lower) sheet. 
On the other hand, for $\sign=-1$, cuts are made in $[-\im\lambda_0,\im\lambda_0]$ 
(see Fig.\ref{fig:contour-}). 
It is required that $\textrm{Im}\ \zeta\ \textrm{Im}\ \lambda>0$ 
($\textrm{Im}\ \zeta\ \text{Im}\ \lambda<0$) on the upper (lower) sheet. 
The Jost functions satisfy the following Volterra-type integral equations, 
\begin{align}
\bm{\phi}^\pm(\lambda,\zeta;x)=J(\zeta x)\left(
I+\int_{\pm\infty}^x \d y J(\zeta y)^{-1}
W \bm{\phi}^\pm (y)\right).
\end{align}
Suppose that 
\begin{align}
\int_{-\infty}^\infty \big| (W_{ab})_{ij} \big| \d x <\infty ,
\end{align}
for all $a, b=1,2$ and $i, j=1,\cdots,l$. 
We may have the Neumann series solution 
\begin{align}
\bm{\phi}^\pm(\lambda,\zeta;x)J(\zeta x)^{-1}&=\sum_{n=0}^\infty
\int_{\pm\infty}^x \!\!\!\d y_1 \int_{\pm\infty}^{y_1} \!\!\!\d y_2 \cdots
\int_{\pm\infty}^{y_{n-1}} \!\!\!\d y_n \ G(y_1)\cdots G(y_n),
\nonumber\\
&\equiv\bm{T}^{<(>)}\e^{\int_{\pm\infty}^x G(y)\d y},
\end{align}
where $G(y)=J\left(\zeta (y-x)\right)^{-1}WJ\left(\zeta (y-x)\right)$ 
and $\bm{T}^{<(>)}$ denotes the time-ordered product.
Examining the convergence of the Neumann series and its derivatives, 
it is found that $\phi_1^-(\lambda,\zeta;x)\e^{\smalli\zeta x}$,
$\phi_2^+(\lambda,\zeta;x)\e^{-\smalli\zeta x}$ are bounded and analytic
in the region where $\textrm{Im}\ \zeta>0$, and 
$\phi_1^+(\lambda,\zeta;x)\e^{\smalli\zeta x}$,
$\phi_2^-(\lambda,\zeta;x)\e^{-\smalli\zeta x}$ are bounded and analytic 
in the region where $\textrm{Im}\ \zeta<0$. 
Relations (\ref{eq:Azeta+}) and (\ref{eq:Abarzeta+})
show that $A(\lambda,\zeta)$ ($\bar{A}(\lambda,\zeta)$) is
analytic in the region where $\textrm{Im}\ \zeta>0$ ($\textrm{Im}\ \zeta<0$). 
We also have the asymptotic behaviors of the Jost functions and the scattering data 
as $\lambda,\zeta\to\infty$ from the asymptotics of $W$: 
\begin{subequations}
\begin{align}
W_{11}&=
\begin{cases}
	-\widetilde{Q}^{-1}\widetilde{Q}_x+\mathcal{O}(1/|\lambda|),
	& \mbox{if } \zeta\simeq\lambda,\\
	\mathcal{O}(1/|\lambda|),
	& \mbox{if } \zeta\simeq -\lambda,
\end{cases}
\\
W_{12}&=\mathcal{O}(1),\\
W_{21}&=\mathcal{O}(1),\\
W_{22}&=
\begin{cases}
	\widetilde{Q}_x\widetilde{Q}^{-1}+\mathcal{O}(1/|\lambda|),
	& \mbox{if } \zeta\simeq\lambda,\\
	\mathcal{O}(1/|\lambda|),
	& \mbox{if } \zeta\simeq -\lambda.
\end{cases}
\end{align}
\end{subequations}
In calculating Neumann series, the following formulae are useful:
\begin{align}
&\bm{T}^> \e^{-\int_a^b \widetilde{Q}^{-1}\widetilde{Q}_x \d x}
=\widetilde{Q}^{-1}(b)\widetilde{Q}(a),
\label{eq:Tprodformula11}\\
&\bm{T}^< \e^{\int_a^b \widetilde{Q}^{-1}\widetilde{Q}_x \d x}
=\widetilde{Q}^{-1}(a)\widetilde{Q}(b),
\label{eq:Tprodformula21}
\end{align}
for $a<b$. 
The proof is as follows. Write the RHS of (\ref{eq:Tprodformula11}) as $X(a)$. 
Differentiating $X(a)$ gives 
$X_a=X\widetilde{Q}^{-1}\widetilde{Q}_a$. Solving this equation, we get 
$X(a)=A\widetilde{Q}(a)$. 
Since $X(b)=I$, we find $A=\widetilde{Q}^{-1}(b)$, and we arrive at the 
desired formula. We get (\ref{eq:Tprodformula21}) just in the same way. 
Consequently, when $\zeta\simeq\lambda$, we have the asymptotics 
\begin{subequations}
\label{eq:asympt-jostAB1}
\begin{align}
\bm{\phi}^\pm(\lambda,\zeta ;x)J(\zeta x)^{-1}&=\left[
\begin{array}{cc}
\lambda_0^{-2}\widetilde{Q}^\dagger(x)Q_\pm & O\\
O & \lambda_0^{-2}\widetilde{Q}(x)Q^\dagger_\pm
\end{array}
\right] +\mathcal{O}(1/|\lambda|),\\
S(\lambda,\zeta)&=\left[
\begin{array}{cc}
\lambda_0^{-2}Q^\dagger_+ Q_- & O\\
O & \lambda_0^{-2}Q_+ Q_-^\dagger
\end{array}
\right] +\mathcal{O}(1/|\lambda|).
\end{align}
\end{subequations}
On the other hand, when $\zeta\simeq -\lambda$, 
\begin{subequations}
\label{eq:asympt-jostAB2}
\begin{align}
\bm{\phi}^\pm(\lambda,\zeta ;x)J(\zeta x)^{-1}&=I+\mathcal{O}(1/|\lambda|),\\
S(\lambda,\zeta)&=I+\mathcal{O}(1/|\lambda|).
\end{align}
\end{subequations}
Furthermore, one can also show
\begin{align}
\label{eq:asympt-Imzeta+}
&\psi_1^-(\lambda,\zeta ;x)\e^{\im\zeta x}[A(\lambda,\zeta)]^{-1}
-T^+(\lambda,\zeta)\left[
\begin{array}{c} I \\ O \end{array}
\right]=\mathcal{O}(1), \qquad \mathrm{Im}\ \zeta>0, \\
&\psi_2^-(\lambda,\zeta ;x)\e^{-\im\zeta x}[\bar{A}(\lambda,\zeta)]^{-1}
-T^+(\lambda,\zeta)\left[
\begin{array}{c} O \\ I \end{array}
\right]=\mathcal{O}(1), \qquad \mathrm{Im}\ \zeta<0.
\end{align}
Note that, however, when $\lambda=\pm\sqrt{\sign}\lambda_0$ the scattering data may have a singularity of the order $\mathcal{O}(1/\zeta)$ which can be seen from eqs.~(\ref{eq:wronskis+}), while the eigenfunctions are well-defined at those branch points in the general situation~\cite{FadTak}.
It should be remarked that the introduction of $\widetilde{Q}(x)$ is irrelevant in the analysis of the original eigenvalue problem (\ref{eq:mtxeigenU}) and the inverse problem discussed in the subsequent section. 
%\begin{subequations}
%\label{eq:asympt-jostAB}
%\begin{align}
%\bm{\psi}^\pm (\lambda,\zeta;x)J(\zeta x)^{-1}&=\left[
%\begin{array}{cc}
%-\im Q_\pm & O\\
%O & -\im\sign Q_\pm^\dagger
%\end{array}
%\right] +\mathcal{O}(1/|\lambda|),\\
%A(\lambda,\zeta)&=\frac{1}{\lambda_0^2}Q_+^\dagger Q_- 
%+\mathcal{O}(1/|\lambda|),\\
%\bar{A}(\lambda,\zeta)&=\frac{1}{\lambda_0^2}Q_+ Q_-^\dagger 
%+\mathcal{O}(1/|\lambda|),\\
%B(\lambda,\zeta)&=\mathcal{O}(1/|\lambda|),\\
%\bar{B}(\lambda,\zeta)&=\mathcal{O}(1/|\lambda|).
%\end{align}
%\end{subequations}

\subsection{Inverse problem}
\label{sec:inverse}
In this subsection,
we derive the Gel'fand--Levitan--Marchenko equations
which give the solution of the inverse problem,
by use of the Jost functions on the complex Riemann surface.

We assume that the Jost functions (\ref{eq:jost+}) are expressed as
\begin{equation}
\label{eq:intjost+}
\bm{\psi}^\pm(\lambda,\zeta;x)=T^\pm(\lambda,\zeta)J(\zeta x)
+\int_x^{\pm\infty}\mathcal{K}(x,s)T^\pm(\lambda,\zeta)J(\zeta s)\d s,
\end{equation}
where the kernel matrix is 
\begin{equation}
\mathcal{K}(x,s)=\left[\begin{array}{cc}
\mathcal{K}_{11}(x,s)&\mathcal{K}_{12}(x,s)\\
\mathcal{K}_{21}(x,s)&\mathcal{K}_{22}(x,s)
\end{array}\right],
\end{equation}
with $\mathcal{K}_{ij}(x,s)$ being $l\times l$ matrix functions.
Substituting the expression (\ref{eq:intjost+}) into the eigenvalue
problem (\ref{eq:mtxeigenU}), after some calculations,
we obtain a linear differential equation for the kernel matrix $\mathcal{K}(x,s)$,
\begin{align}
\label{eq:linearsys+}
\partial_x\mathcal{K}(x,s)+&
(\sigma^z\otimes I)\partial_s\mathcal{K}(x,s)(\sigma^z\otimes I)\nonumber\\
&+(\sigma^z\otimes I)\mathcal{K}(x,s)(\sigma^z\otimes I)
\left[U^\pm(\lambda,\zeta)+\im\lambda(\sigma^z\otimes I)\right]\nonumber\\
&-\left[U(\lambda,\zeta;x)+\im\lambda(\sigma^z\otimes I)\right]
\mathcal{K}(x,s)=O,
\end{align}	
with boundary conditions,
\begin{subequations}
\label{eq:kernelBC}
\begin{align}
\label{eq:kernelBC1}
2\mathcal{K}_{12}(x,x)&=Q_+ -Q(x),\\
\label{eq:kernelBC2}
2\mathcal{K}_{21}(x,x)&=\sign(Q_+^\dagger-Q^\dagger(x)),\\
\mathcal{K}_{ij}(x,s) &\to O\quad\mbox{as}\quad s\to \pm\infty.
\end{align}
\end{subequations}
This type of a linear system is known as the Marchenko equations and can be uniquely solved, which guarantees the existence of the expression (\ref{eq:intjost+}).

\begin{figure}[h]
\begin{center}
\includegraphics{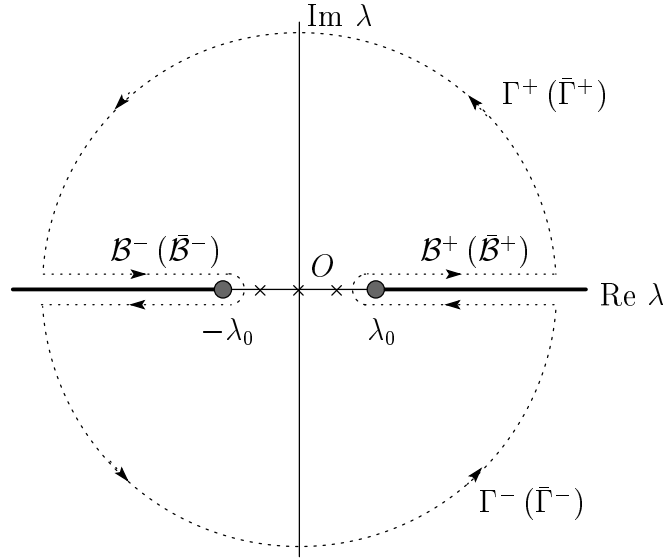}
  \caption[Cuts and integral contours
	on the upper sheet ($\textrm{Im}\ \zeta>0$) of $\lambda$ plane.]
	{Cuts, bold lines, and integral contours, dotted lines, 
	on the upper (lower) sheet of the Riemann surface of $\lambda$ plane
	for the self-defocusing case, $\sign=+1$.
	The cross denotes the zero for $\det A$ ($\det\bar{A}$) on the upper (lower) sheet.}
  \label{fig:contour+}
\end{center}
\end{figure}

First, we concentrate on $\sign=+1$ case. 
Before going on, we mention several facts.
For the characteristic roots (\ref{eq:root}), 
we have introduced two branch cuts on the real axis,
$(-\infty,-\lambda_0]$ and $[\lambda_0,\infty)$,
as shown in Fig.~\ref{fig:contour+}.
We define contour paths $\mathcal C$ enclosing a region in the upper sheet
($\textrm{Im}\,\,\zeta>0$) of the Riemann surface of $\lambda$
and $\bar{\mathcal C}$ enclosing a region in the lower sheet
($\textrm{Im}\,\,\zeta<0$) as (see Fig.~\ref{fig:contour+}),
\begin{subequations}
\begin{align}
\mathcal C=&\Gamma+\mathcal B,\qquad\quad\bar{\mathcal C}
=\bar{\Gamma}+\bar{\mathcal B},\\
\Gamma=&\Gamma^++\Gamma^-,
\quad\,\,\,\,\bar{\Gamma}=\bar{\Gamma}^++\bar{\Gamma}^-,\\
\mathcal B=&\mathcal B^++\mathcal B^-,\quad\,\,\,\bar{\mathcal B}
=\bar{\mathcal B}^++\bar{\mathcal B}^-,
\end{align}
\end{subequations}
where the superscripts $+$ and $-$ denote 
the part of the path $\Gamma$ ($\bar{\Gamma}$) which exists
in the upper and lower half plane of each sheet, respectively,
and the part of the path $\mathcal B$ ($\bar{\mathcal B}$) which exists in
the right and left half plane of each sheet, respectively. 
The radius of $\Gamma$ ($\bar{\Gamma}$) is assumed to be large enough for 
$\mathcal{C}$ ($\bar{\mathcal{C}}$) to enclose all the zeros of 
$\det A$ ($\det\bar{A}$). 
As noticing in Sec. \ref{sec:direct}, if $(\lambda_j,\zeta_j)$ is a zero 
of $\det A$, it holds that $\lambda_j\in\mathbb{R}$ and $\zeta_j\in\im\mathbb{R}$.  
Correspondingly,  $(\lambda_j,-\zeta_j)$ is a zero of $\det\bar{A}$. 
Suppose here that $\det A$ and $\det\bar{A}$ have $N$ zeros, respectively. 
%As noticing in Sec.\ref{sec:direct}, since a pair of zeros of $\det A$, say 
%$(\lambda_j,\zeta_j)$ and $(\lambda_j^*,-\zeta_j^*)$, 
%lie on the region where $\mathrm{Im}\ \zeta>0$, 
%it is necessary that $\lambda_j=\lambda_j^*\in\mathbb{R}$ and 
%$\zeta_j=-\zeta_j^*\in\im\mathbb{R}$. 
Along the branch cut in the upper sheet, one can show that
\begin{align}
&\int_{\mathcal B}\d \lambda\frac{\lambda}{\zeta}
\e^{\smalli\zeta z}=4\pi\delta(z),
\label{eq:intB1}\\
&\int_{\mathcal B}\d \lambda\e^{\smalli\zeta z}=
\int_{\mathcal B}\d \lambda\frac{\e^{\smalli\zeta z}}{\zeta}=0,
\label{eq:intB2}
\end{align}
where $z$ is real and $\delta(z)$ denotes the delta function.
Using these formulae, we obtain 
\begin{equation}
\label{eq:intformula}
\frac{1}{4\pi}\int_{\mathcal B}\d \lambda\frac{\e^{\smalli\zeta z}}{\zeta}
T^\pm(\lambda,\zeta)=\delta(z)(\sigma^x\otimes I).
\end{equation}
In the lower sheet, integral formulae (\ref{eq:intB1}), (\ref{eq:intB2}) and (\ref{eq:intformula}) 
hold with replacing $\mathcal{B}\to\bar{\mathcal{B}}$, $\zeta\to-\zeta$. 

Going back to eqs.~(\ref{eq:intjost+}), we explicitly write as 
\begin{subequations}
\label{eq:intrep}
\begin{align}
\label{eq:intrep1}
\psi^+_1(\lambda,\zeta;x)&=T^+(\lambda,\zeta)\e^{-\smalli\zeta x}
\left[\begin{array}{c}I\\O\end{array}\right]
+\int_x^{\infty}\d s\mathcal{K}(x,s)
T^+(\lambda,\zeta)\e^{-\smalli\zeta s}
\left[\begin{array}{c}I\\O\end{array}\right],\\
\label{eq:intrep2}
\psi^+_2(\lambda,\zeta;x)&=T^+(\lambda,\zeta)\e^{\smalli\zeta x}
\left[\begin{array}{c}O\\I\end{array}\right]
+\int_x^{\infty}\d s\mathcal{K}(x,s)
T^+(\lambda,\zeta)\e^{\smalli\zeta s}
\left[\begin{array}{c}O\\I\end{array}\right].
\end{align}
\end{subequations}
We rewrite eq.~(\ref{eq:defscatmtx}) into 
\begin{subequations}
\begin{align}
\label{eq:phiainv+}
\psi^-_1(\lambda,\zeta;x)[A(\lambda,\zeta)]^{-1}
=&\psi^+_1(\lambda,\zeta;x)+
\psi^+_2(\lambda,\zeta;x)B(\lambda,\zeta)[A(\lambda,\zeta)]^{-1},\\
\label{eq:phibarainv+}
\psi^-_2(\lambda,\zeta;x)[\bar{A}(\lambda,\zeta)]^{-1}
=&\psi^+_2(\lambda,\zeta;x)
+\psi^+_1(\lambda,\zeta;x)\bar{B}(\lambda,\zeta)
[\bar{A}(\lambda,\zeta)]^{-1}.
\end{align}
\end{subequations}
Substituting eq.~(\ref{eq:intrep1}) into eq.~(\ref{eq:phiainv+}),
we have a relation,
\begin{align}
\label{eq:phi1-}
\psi^-_1(\lambda,\zeta;x)[A(\lambda,\zeta)]^{-1}
&-T^+(\lambda,\zeta)\e^{-\smalli\zeta x}
\left[\begin{array}{c}I\\O\end{array}\right]\nonumber\\
=&\int_x^{\infty}\d s\mathcal{K}(x,s)
	T^+(\lambda,\zeta)\e^{-\smalli\zeta s}
	\left[\begin{array}{c}I\\O\end{array}\right]\nonumber\\
&+\left\{T^+(\lambda,\zeta)\e^{\smalli\zeta x}
\left[\begin{array}{c}O\\I\end{array}\right]\right.\nonumber\\
&\left.+\int_x^{\infty}\d s\mathcal{K}(x,s)
T^+(\lambda,\zeta)\e^{\smalli\zeta s}
\left[\begin{array}{c}O\\I\end{array}\right]\right\}
B(\lambda,\zeta)[A(\lambda,\zeta)]^{-1}.
\end{align}
When we multiply
\begin{equation}
\label{eq:multiplier}
\frac{\e^{\smalli \zeta y}}{4\pi\zeta}
\quad\quad(y>x)
\end{equation}
on the both sides of eq.~(\ref{eq:phi1-}),
the left hand side of eq.~(\ref{eq:phi1-}) becomes analytic on the 
upper sheet of the Riemann surface ($\textrm{Im}\,\,\zeta>0$),
with the exception of the 
points $\lambda_j$, at which it has simple poles.
Here we have assumed that $1/\det A(\lambda,\zeta)$ has $N$
isolated simple poles $\{\lambda_1,\lambda_2,\dots,\lambda_N\}$
in the upper sheet.
From %eqs.~(\ref{eq:T+-}) and (\ref{eq:asympt-jostAB2}), 
eq.~(\ref{eq:asympt-Imzeta+}), 
we can show that 
as $|\lambda|\to\infty$, the left hand side of the resultant equation
behaves like $\exp[-\textrm{Im}\,\,\zeta(y-x)]\mathcal{O}(1/|\lambda|)$.
We integrate the relation (\ref{eq:phi1-}) multiplied by 
eq.~(\ref{eq:multiplier}) along the contour $\mathcal B$.
In the integration of the left hand side, the contour can be closed
through infinity, i.e., $\mathcal C=\mathcal B+\Gamma$,
so that the integral of the left hand side is 
equal to the sum of the residues at $\lambda=\lambda_j$,
\begin{align}
\label{eq:phi1-lint}
\frac{1}{4\pi}\int_{\mathcal C}\d \lambda\frac{\e^{\smalli\zeta(y-x)}}{\zeta}
&\left\{\psi^-_1(\lambda,\zeta;x)\e^{\smalli\zeta x}
[A(\lambda,\zeta)]^{-1}
-T^+(\lambda,\zeta)\left[\begin{array}{c}I\\O\end{array}\right]
\right\}
\nonumber\\
&=\frac{\im}{2}\sum_{j=1}^N\frac{\e^{\smalli\zeta_jy}}{\zeta_j}
\psi^-_1(\lambda_j,\zeta_j;x)
\frac{\widetilde{A}(\lambda_j,\zeta_j)}{(\det A)'(\lambda_j,\zeta_j)}\nonumber\\
&=\im\sum_{j=1}^N \e^{\smalli\zeta_jy}\psi^+_2(\lambda_j,\zeta_j;x) \Pi_j,
\end{align}
where $\widetilde{A}$ is the cofactor matrix of $A$ and 
$\Pi_j$ is the residue matrix at $\lambda=\lambda_j$ defined by
\begin{align}
\label{eq:residue mtx}
\Pi_j=\mathop{\mathrm{Res}}_{\lambda=\lambda_j, \zeta=\zeta_j} 
\left[ \frac{1}{2\zeta} B(\lambda,\zeta)[A(\lambda,\zeta)]^{-1}\right].
\end{align}
In the second equality of eq.~(\ref{eq:phi1-lint}), we have used a relation
deduced from eq.~(\ref{eq:phiainv+}) at $(\lambda,\zeta)=(\lambda_j,\zeta_j)$, 
which means that 
$\psi_1^-(\lambda_j,\zeta_j;x)$ is proportional to $\psi_2^+(\lambda_j,\zeta_j;x)$. 
The integral in the right hand side of eq.~(\ref{eq:phi1-lint}) is transformed into
\begin{align}
\label{eq:phi1-rint}
\mathcal{K}(x,y)\left[\begin{array}{c}O\\I\end{array}\right]
+\mathcal{F}_c(x+y)\left[\begin{array}{c}O\\I\end{array}\right]
+\int_x^{\infty}\d s\mathcal{K}(x,s)\mathcal{F}_c(s+y)
\left[\begin{array}{c}O\\I\end{array}\right],
\end{align}
where
\begin{align}
\mathcal{F}_c(z)=&
\frac{1}{4\pi}\int_{\mathcal B}\d\lambda\frac{\e^{\smalli \zeta z}}{\zeta}
T^+(\lambda,\zeta)\rho(\lambda,\zeta),\\
\rho(\lambda,\zeta)=&B(\lambda,\zeta)[A(\lambda,\zeta)]^{-1}.
\end{align}
From the definition of $T^+$ in eq.~(\ref{eq:T+-}), we obtain
the following form,
\begin{equation}
\label{eq:Fc}
\mathcal{F}_c(z)\left[\begin{array}{c}O\\I\end{array}\right]
=\left[\begin{array}{c}\im \mathcal{F}'_{1c}(z)+\mathcal{F}_{2c}(z)\\
\im \sign Q_+^\dagger\mathcal{F}_{1c}(z)
\end{array}\right],
\end{equation}
where $\mathcal{F}'_{1c}(z)\equiv\d\mathcal{F}_{1c}(z)/\d z$, and
\begin{subequations}
\begin{align}
\mathcal{F}_{1c}(z)=&
\frac{1}{4\pi}\int_{-\infty}^\infty\d\xi
\frac{\e^{\smalli\xi z}}{\lambda}\left[\rho(\lambda,\xi)
-\rho(-\lambda,\xi)\right],\\
\mathcal{F}_{2c}(z)=&
\frac{1}{4\pi}\int_{-\infty}^\infty\d\xi
\e^{\smalli\xi z}\left[\rho(\lambda,\xi)+\rho(-\lambda,\xi)\right].
\end{align}
\end{subequations}
Using the integral representation (\ref{eq:intrep2}) in 
eq.~(\ref{eq:phi1-lint}), we rewrite eq.~(\ref{eq:phi1-lint})
into the form (\ref{eq:Fc}) as follows, 
\begin{equation}
\label{eq:Fd}
\mathcal{F}_d(z)\left[\begin{array}{c}O\\I\end{array}\right]
=\left[\begin{array}{c}\im \mathcal{F}'_{1d}(z)+\mathcal{F}_{2d}(z)\\
\im \sign Q_+^\dagger\mathcal{F}_{1d}(z)
\end{array}\right],
\end{equation}
where 
\begin{subequations}
\begin{align}
\mathcal{F}_{1d}(z)=&
\sum_{j=1}^N\im\Pi_j\e^{\smalli\zeta_jz},\\
\mathcal{F}_{2d}(z)=&
\sum_{j=1}^N\im\lambda_j\Pi_j\e^{\smalli\zeta_jz}.
\end{align}
\end{subequations}
Combining eqs.~(\ref{eq:phi1-rint}) and (\ref{eq:Fd}), we finally
obtain the integral equation: %Gel'fand--Levitan--Marchenko equation:
\begin{align}
\label{eq:glmeq1+}
\mathcal{K}(x,y)\left[\begin{array}{c}O\\I\end{array}\right]+
\mathcal{F}(x+y)\left[\begin{array}{c}O\\I\end{array}\right]
+\int^\infty_x\d s\mathcal{K}(x,s)\mathcal{F}(s+y)
\left[\begin{array}{c}O\\I\end{array}\right]=
\left[\begin{array}{c}O\\O\end{array}\right]\quad\quad
(y>x).
\end{align}
Here
\begin{equation}
\mathcal{F}(z)=\mathcal{F}_c(z)-\mathcal{F}_d(z).
\end{equation}
Following the same procedure using eqs.~(\ref{eq:intrep2}) and (\ref{eq:phibarainv+}), 
we obtain 
\begin{align}
\label{eq:glmeq2+}
\mathcal{K}(x,y)\left[\begin{array}{c}I\\O\end{array}\right]+
\bar{\mathcal{F}}(x+y)\left[\begin{array}{c}I\\O\end{array}\right]
+\int^\infty_x\d s\mathcal{K}(x,s)\bar{\mathcal{F}}(s+y)
\left[\begin{array}{c}I\\O\end{array}\right]=
\left[\begin{array}{c}O\\O\end{array}\right]\quad\quad
(y>x),
\end{align}
where
\begin{align}
\bar{\mathcal{F}}(z)=
&\frac{1}{4\pi}\int_{\bar{\Gamma}} \d\lambda
\frac{\e^{-\smalli\zeta z}}{\zeta}T^+(\lambda,\zeta)
\bar{\rho}(\lambda,\zeta),\\
\bar{\rho}(\lambda,\zeta)
=&\bar{B}(\lambda,\zeta)[\bar{A}(\lambda,\zeta)]^{-1}.
\end{align}
The residue matrix at the zero of $\det\bar{A}$, say, 
$(\lambda,\zeta)=(\bar{\lambda}_j,\bar{\zeta}_j)$ 
is defined by 
\begin{align}
\bar{\Pi}_j=\mathop{\mathrm{Res}}_{\lambda=\bar{\lambda}_j,\zeta=\bar{\zeta}_j} 
\left[ -\frac{1}{2\zeta} \bar{B}(\lambda,\zeta)[\bar{A}(\lambda,\zeta)]^{-1}\right].
\label{eq:barPij}
\end{align}
Integral equations (\ref{eq:glmeq1+}) and (\ref{eq:glmeq2+}) are called 
the Gel'fand--Levitan--Marchenko equations. 
Solving these equations as to the kernel $\mathcal{K}(x,y)$ for given $\mathcal{F}(z)$
and $\bar{\mathcal{F}}(z)$, we can obtain the potential matrix $Q$ 
through the relations (\ref{eq:kernelBC}). 

\begin{figure}[h]
\begin{center}
\includegraphics{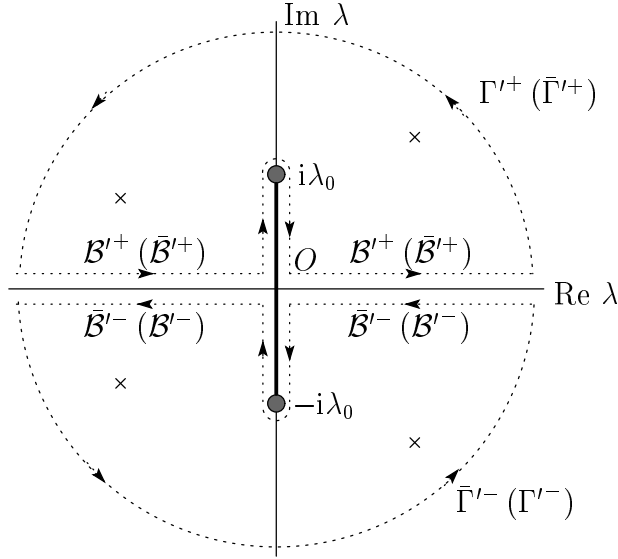}
\end{center}
\caption{A cut, bold line, and integral contours, dotted lines, 
	on the upper (lower) sheet of the Riemann surface of $\lambda$ plane
	for the self-focusing case, $\sign=-1$. 
	The cross denotes the zero for $\det A$ and $\det\bar{A}$ on both sheets.}
\label{fig:contour-}
\end{figure}

Let us move on to $\sign=-1$ case. 
Integral contours are depicted in Fig.\ref{fig:contour-}. 
The branch cut in the Riemann surface is made along $[-\im\lambda_0,\im\lambda_0]$. 
Contours playing the same role as in the previous case are named with the same letter 
with a prime $'$. 
We define a contour path $\mathcal{C}'$ ($\bar{\mathcal{C}}'$) 
enclosing a region $\mathrm{Im}\ \zeta>0$ ($\mathrm{Im}\ \zeta<0$) as, 
\begin{subequations}
\begin{align}
\mathcal{C}'=&\Gamma'+\mathcal{B}',\qquad\quad\bar{\mathcal C}'
=\bar{\Gamma}'+\bar{\mathcal B}',\\
\Gamma'=&\Gamma'^+ +\Gamma'^-,
\quad\,\,\,\,\bar{\Gamma}'=\bar{\Gamma}'^+ +\bar{\Gamma}'^-,\\
\mathcal{B}'=&\mathcal{B}'^++\mathcal{B}'^-,\quad\,\,\,\bar{\mathcal B}'
=\bar{\mathcal B}'^++\bar{\mathcal B}'^-,
\end{align}
\end{subequations}
where the superscripts $+$ and $-$ denote the part of paths in the upper and lower 
half plane of each sheet. 
The contours $\mathcal{B}'^\pm$ and $\bar{\mathcal{B}}'^\pm$ are 
along both the real axis and the branch cut. 
The radius of $\Gamma'$ ($\bar{\Gamma}'$) is assumed 
to be large enough for $\mathcal{C}'$ ($\bar{\mathcal{C}}'$) to enclose 
all the zeros of $\det A$ ($\det\bar{A}$). 
In contrast to the previous case, the zero of $\det A$ is complex and makes a pair. 
Assume that there are $N=2M$ zeros for $\det A$ and label them such that 
$\lambda_{2k}=\lambda_{2k-1}^*$, $\zeta_{2k}=-\zeta_{2k-1}^*$ for $k=1,\cdots,M$. 
Correspondingly, we have the zeros of $\det\bar{A}$ such that 
$\{(\lambda_1,-\zeta_1),(\lambda_1^*,\zeta_1^*),\cdots,(\lambda_M,-\zeta_M),(\lambda_M^*,\zeta_M^*)\}$. 
The derivation of the integral equation goes in parallel 
to that for the case $\sign=+1$. 
As a consequence, we arrive at just the same equations 
(\ref{eq:glmeq1+}) and (\ref{eq:glmeq2+}).

We remark on the properties of the matrices $\Pi_j$ and $\bar{\Pi}_j$. 
First, the determinant is zero. 
This is proved directly as
\begin{align}
\det\Pi_j&\propto \frac{\det B(\det A)^{l-1}}{[(\det A)']^{l}}(\lambda_j,\zeta_j)
\nonumber\\
&=0,
\end{align}
where we have used $\det A(\lambda_j,\zeta_j)=0$. 
Similarly, we have $\det\bar{\Pi}_j=0$. 
Second, we have the following relations:
\begin{subequations}
\label{eq:relPi}
\begin{align}
\Pi_j=\Pi_j^\dagger=-\bar{\Pi}_j&\qquad(\sign=+1),\\
\Pi_{2j-1}=\Pi_{2j}^\dagger=\bar{\Pi}_{2j-1}=\bar{\Pi}_{2j}^\dagger&\qquad(\sign=-1).
\end{align}
\end{subequations}
These are proved as follows. From eqs.~(\ref{eq:ABinv+}), (\ref{eq:parity1}), 
and (\ref{eq:parity2}), we obtain
\begin{align}
&\bar{B}\bar{A}^{-1}(\lambda,\zeta)=\sign [(BA^{-1})(\lambda^*,\zeta^*)]^\dagger,
\label{eq:rel1}\\
&\bar{B}\bar{A}^{-1}(\lambda,\zeta)=-\sign BA^{-1}(\lambda,-\zeta).
\label{eq:rel2}
\end{align}
For example, we demonstrate $\bar{\Pi}_j=-\Pi_j^\dagger$ for $\sign=+1$. 
Substituting eq.~(\ref{eq:rel1}) into the definition of $\bar{\Pi}_j$ (\ref{eq:barPij}) we have
\begin{align}
\bar{\Pi}_j&=
\mathop{\mathrm{Res}}_{\lambda=\lambda_j,\zeta=-\zeta_j} 
\left[-\frac{1}{2\zeta}[(BA^{-1})(\lambda^*,\zeta^*)]^\dagger\right]\nonumber\\
&=-\left\{\mathop{\mathrm{Res}}_{\lambda^*=\lambda_j^*,\zeta^*=-\zeta_j^*} 
\left[\frac{1}{2\zeta^*}(BA^{-1})(\lambda^*,\zeta^*)\right]\right\}^\dagger\nonumber\\
&=-\left\{\mathop{\mathrm{Res}}_{\lambda=\lambda_j,\zeta=\zeta_j} 
\left[\frac{1}{2\zeta}(BA^{-1})(\lambda,\zeta)\right]\right\}^\dagger\nonumber\\
&=-\Pi_j^\dagger,
\end{align}
where in the third equality we have replaced the dummy variables as $\lambda^*\to\lambda$, $\zeta^*\to\zeta$ and used $\lambda^*_j=\lambda_j$, $\zeta_j^*=-\zeta_j$.
The rests are obtained similarly.

\subsection{Time dependence of the scattering data}
Next, we consider the time dependence of the scattering data.
Under the nonvanishing boundary conditions (\ref{eq:constBC}), 
the asymptotic forms of the Lax matrix $V$ are given by
\begin{equation}
\label{eq:Vasympt}
V\to V^\pm= 2(\lambda-k)\left[
\begin{array}{cc}
-2\im\lambda I&Q_\pm\\
\sign Q_\pm^\dagger&2\im\lambda I
\end{array}\right]\quad \mbox{as }\,\, x\to \pm\infty.
\end{equation}
Operating (\ref{eq:Vasympt}) on the asymptotic forms of
the Jost functions (\ref{eq:jostasympt+}) gives
\begin{subequations}
\begin{align}
V^\pm T^\pm(\lambda,\zeta)\left[\begin{array}{c}I \\O
\end{array}\right]\e^{-\smalli \zeta x}=&
-2\im(\lambda-k)[\zeta+\lambda(\sigma^z\otimes I)]
T^\pm(\lambda,\zeta)\left[\begin{array}{c}I \\O
\end{array}\right]\e^{-\smalli \zeta x},\\
V^\pm T^\pm(\lambda,\zeta)\left[\begin{array}{c}O \\I
\end{array}\right]\e^{\smalli \zeta x}=&
2\im(\lambda-k)[\zeta-\lambda(\sigma^z\otimes I)]
T^\pm(\lambda,\zeta)\left[\begin{array}{c}O \\I
\end{array}\right]\e^{\smalli \zeta x}.
\end{align}
\end{subequations}
Then we define the time-dependent Jost functions $\psi^{\pm(t)}_j$, $j=1,2$, as,
\begin{subequations}
\begin{align}
\psi^{\pm(t)}_1&\equiv
\e^{-2\smalli(\lambda-k)[\zeta+\lambda(\sigma^z\otimes I)]t}
\psi^\pm_1\nonumber\\
&\sim \e^{-2\smalli(\lambda-k)[\zeta+\lambda(\sigma^z\otimes I)]t
-\smalli\zeta x}
T^\pm\left[\begin{array}{c}I \\O
\end{array}\right]
\quad \mbox{as }\,\, x\to \pm\infty,\\
\psi^{\pm(t)}_2&\equiv
\e^{2\smalli(\lambda-k)[\zeta-\lambda(\sigma^z\otimes I)]t}
\psi^\pm_2\nonumber\\
&\sim \e^{2\smalli(\lambda-k)[\zeta-\lambda(\sigma^z\otimes I)]t
+\smalli\zeta x}
T^\pm\left[\begin{array}{c}O \\I
\end{array}\right]
\quad \mbox{as }\,\, x\to \pm\infty,
\end{align}
\end{subequations}
which obey, respectively,
\begin{subequations}
\begin{align}
&\frac{\partial\psi^{\pm(t)}_1}{\partial t}=V\psi^{\pm(t)}_1,\\
&\frac{\partial\psi^{\pm(t)}_2}{\partial t}=V\psi^{\pm(t)}_2.
\end{align}
\end{subequations}
These relations give
\begin{subequations}
\label{eq:tdepphi+}
\begin{align}
&\frac{\partial\psi^-_1}{\partial t}=
\{V+2\im(\lambda-k)[\zeta+\lambda(\sigma^z\otimes I)]\}\psi^-_1,
\\
&\frac{\partial\psi^-_2}{\partial t}=
\{V-2\im(\lambda-k)[\zeta-\lambda(\sigma^z\otimes I)]\}\psi^-_2.
\end{align}
\end{subequations}
Substituting the definitions of the scattering data
(\ref{eq:defscatmtx}):
\begin{subequations}
\begin{align}
\psi^-_1(\lambda,\zeta;x,t)=
&\psi^+_1(\lambda,\zeta;x,t)A(\lambda,\zeta;t)
+\psi^+_2(\lambda,\zeta;x,t)B(\lambda,\zeta;t),\\
\psi^-_2(\lambda,\zeta;x,t)=
&\psi^+_1(\lambda,\zeta;x,t)\bar{B}(\lambda,\zeta;t)
+\psi^+_2(\lambda,\zeta;x,t)\bar{A}(\lambda,\zeta;t),
\end{align}
\end{subequations}
into eqs.~(\ref{eq:tdepphi+}), and then taking the limit $x\to\infty$,
we find the time dependence of the scattering data as follows:
\begin{subequations}
\label{eq:scatdat1+}
\begin{align}
\label{eq:scatdat1a+}
A(\lambda,\zeta;t)=&A(\lambda,\zeta;0),\\
B(\lambda,\zeta;t)=&B(\lambda,\zeta;0)\e^{4\smalli \zeta(\lambda-k) t},\\
\Pi_j(t)=&\Pi_j(0)\e^{4\smalli \zeta_j(\lambda_j-k) t}.
\end{align}
\end{subequations}
Using eqs.~(\ref{eq:scatdat1+}), %and (\ref{eq:scatdat2+}),
we obtain explicit time dependence of
$\mathcal{F}_1(z,t)$ and $\mathcal{F}_2(z,t)$,
\begin{subequations}
\label{eq:tdF+}
\begin{align}
\mathcal{F}_1(z,t)=&\frac{1}{4\pi}\int^\infty_{-\infty}\d \xi
\frac{\e^{\smalli \xi z+4\smalli\xi(\lambda-k) t}}{\lambda}
\left[\rho(\lambda,\xi;0)-\rho(-\lambda,\xi;0)\right]-
\sum^N_{j=1}\im\Pi_j(0)\e^{\smalli\zeta_j z+4\smalli\zeta_j(\lambda_j-k)t},\\
\mathcal{F}_2(z,t)=&\frac{1}{4\pi}\int^\infty_{-\infty}\d \xi
\e^{\smalli \xi z+4\smalli\xi(\lambda-k) t}
\left[\rho(\lambda,\xi;0)+\rho(-\lambda,\xi;0)\right]-
\sum^N_{j=1}\im\lambda_j\Pi_j(0)
\e^{\smalli\zeta_j z+4\smalli\zeta_j(\lambda_j-k)t}.
\end{align}
\end{subequations}
We have similar time dependence for $\bar{\mathcal{F}}_1(z,t)$ and 
$\bar{\mathcal{F}}_2(z,t)$ with $\bar\rho$ and $\bar{\Pi}_j$. 

The procedure of the ISM for 
the initial value problem of the matrix NLS equation (\ref{eq:genNLS}) 
is summarized as follows. First, we solve the eigenvalue problem (\ref{eq:mtxeigenU}) 
for the initial value $Q(x,0)$ and obtain the scattering data 
at $t=0$ (direct problem). 
Then, with the time-dependent scattering data (\ref{eq:scatdat1+}), 
we solve the Gel'fand--Levitan--Marchenko equations (\ref{eq:glmeq1+}) and 
(\ref{eq:glmeq2+}) to obtain $\mathcal{K}(x,y,t)$ and $Q(x,t)$ (inverse problem). 
This procedure provides the direct proof of the complete integrability of 
the matrix NLS equation (\ref{eq:genNLS}) 
under the nonvanishing boundary conditions (\ref{eq:constBC}).
Further researches are required to establish each step in a rigorous way.

\subsection{Soliton solutions}
We construct soliton solutions of the matrix NLS equation
under the reflection-free condition:
\begin{equation}
B(\lambda,\xi)=\bar{B}(\lambda,\xi)=O, \qquad (\xi\mbox{ : real}),
\end{equation}
whereby the first terms, $\mathcal{F}_c(z,t)$ parts, in eqs.~(\ref{eq:tdF+})
identically vanish,
\begin{align}
\label{eq:reducedF+}
\mathcal{F}_1(z,t)=-\sum^N_{j=1}\im\Pi_j(t)\e^{\smalli\zeta_j z},\qquad
\mathcal{F}_2(z,t)=-\sum^N_{j=1}\im\lambda_j\Pi_j(t)\e^{\smalli\zeta_j z}.
\end{align}
and similarly for $\bar{\mathcal{F}}_1$ and $\bar{\mathcal{F}}_2$. 
Assume the form, 
\begin{align}
&\mathcal{K}_{11}(x,y,t)=
\im\lambda_0\sum_{j=1}^{N} \mathcal{K}_{11}^{(j)}\bar{\Pi}_j(t) \e^{\im\zeta_j (x+y)},\\
&\mathcal{K}_{12}(x,y,t)=
\im\lambda_0\sum_{j=1}^{N} \mathcal{K}_{12}^{(j)}\Pi_j(t) \e^{\im\zeta_j (x+y)}.
\end{align}
Then, the Gel'fand--Levitan--Marchenko equations 
(\ref{eq:glmeq1+}) and (\ref{eq:glmeq2+}) 
are reduced to a set of linear algebraic equations, 
\begin{align}
&\im\lambda_0\mathcal{K}_{12}^{(j)}+\im(\zeta_j-\lambda_j)I
+\sum_{k=1}^{N}\frac{\lambda_0(\zeta_j-\lambda_j)}{\im(\zeta_j+\zeta_k)}
\mathcal{K}_{11}^{(k)}\bar{\Pi}_k(t) \e^{\im\zeta_k x}
+\sum_{k=1}^{N}\frac{-\im\sign\lambda_0^2}{\im(\zeta_j+\zeta_k)}
\mathcal{K}_{12}^{(k)}\Pi_k(t) \e^{\im\zeta_k x}=O,\\
&\im\lambda_0\mathcal{K}_{11}^{(j)}-\lambda_0 I
+\sum_{k=1}^{N}\frac{\im\lambda_0^2}{\im(\zeta_j+\zeta_k)}
\mathcal{K}_{11}^{(k)}\bar{\Pi}_k(t) \e^{\im\zeta_k x}
+\sum_{k=1}^{N}\frac{-\lambda_0(\zeta_j+\lambda_j)}{\im(\zeta_j+\zeta_k)}
\mathcal{K}_{12}^{(k)}\Pi_k(t) \e^{\im\zeta_k x}=O,
\end{align}
for $j=1,\cdots,N$. Those are simplified into 
\begin{align}
&\mathcal{K}_{11}^{(j)}=\frac{\im\lambda_0}{\zeta_j-\lambda_j}\mathcal{K}_{12}^{(j)},\\
&-I=\frac{\lambda_0}{\zeta_j-\lambda_j}\mathcal{K}_{12}^{(j)}
+\sum_{k=1}^{N} 
\frac{\lambda_0^2 \e^{2\im\zeta_k x}}{\im(\zeta_j+\zeta_k)(\zeta_k-\lambda_k)}
\mathcal{K}_{12}^{(k)}\bar{\Pi}_k(t)
-\sum_{k=1}^{N}
\frac{\sign\lambda_0^2\e^{2\im\zeta_k x}}{\im(\zeta_j+\zeta_k)(\zeta_j-\lambda_j)}
\mathcal{K}_{12}^{(k)}\Pi_k(t).
\label{eq:K12}
\end{align}
Replace $\lambda_j\to-\lambda_j$ for notational reason, 
and note the relations (\ref{eq:relPi}) and 
that the time-dependence of $\Pi_j$ and $\bar{\Pi}_j$ is
\begin{align}
\Pi_j(t)=\Pi_j(0)\e^{\chi_j},\qquad
\bar{\Pi}_j(t)=\bar{\Pi}_j(0)\e^{\chi_j},
\end{align}
where we introduce the function 
\begin{align}
\chi_j(x,t)=2\im\zeta_j(x-2(\lambda_j+k)t). 
\end{align}
Then, from eq.~(\ref{eq:K12}) we have a matrix form, 
\begin{align}
\label{eq:glmmtx}
\left(\mathcal{K}_{12}^{(1)}\cdots\mathcal{K}_{12}^{(N)}\right)
\left[
\begin{array}{ccc}
S_{1,1}&\cdots&S_{1,N}\\
\vdots& &\vdots\\
S_{N,1}&\cdots&S_{N,N}
\end{array}
\right]
=-(\overbrace{I\cdots I}^{N}),
\end{align}
where for $i,j=1,\cdots,N$, 
\begin{align}
S_{ij}=\frac{\lambda_0}{\zeta_j+\lambda_j}\delta_{ij}I
-\frac{\sign\lambda_0^2}{\im(\zeta_i+\zeta_j)}
\left(\frac{1}{\zeta_i+\lambda_i}+\frac{1}{\zeta_j+\lambda_j}\right)\Pi_i\e^{\chi_i}.
\end{align}
Solving the linear equations (\ref{eq:glmmtx}) and using eq.~(\ref{eq:kernelBC1}), 
we arrive at the multi-soliton solution
for the modified version of the matrix NLS equation (\ref{eq:mtxNLS+tilde}), 
\begin{align}
Q(x,t)=\lambda_0I+2\im\lambda_0(\overbrace{I\cdots I}^{N})S^{-1}\left[
\begin{array}{c}
\Pi_1\e^{\chi_1}\\
\vdots\\
\Pi_{N}\e^{\chi_{N}}
\end{array}
\right].
\label{eq:Nsoliton}
\end{align}
Before concluding, we summarize about the parameters of the solution. 
For $\sign=+1$, the solution (\ref{eq:Nsoliton}) is the $N$-soliton solution. 
$\lambda_j$ ($j=1,\cdots,N$) is a real constant such that 
$-\lambda_0<\lambda_j<\lambda_0$. 
$\zeta_j=\im(\lambda_0^2-\lambda_j^2)^{1/2}$ is pure imaginary.
$\Pi_j$ is an $l\times l$ Hermitian matrix. 
For $\sign=-1$, the solution (\ref{eq:Nsoliton}) is the $M (=N/2)$-soliton solution. 
$\lambda_j$ and $\zeta_j=(\lambda_j^2+\lambda_0^2)^{1/2}$ ($j=1,\cdots,N$)
are complex constants satisfying 
$\lambda_{2k}=\lambda_{2k-1}^*$ and $\zeta_{2k}=-\zeta_{2k-1}^*$ for $k=1,\cdots,N/2$. 
$\Pi_j$ is an $l\times l$ matrix satisfying $\Pi_{2k-1}=\Pi_{2k}^\dagger$. 

Although, in a strict sense, we should impose that $\det\Pi_j=0$ for all $j$, 
we can relax this condition. 
The reason is that in the limiting case where two distinct 
$\lambda_i$ and $\lambda_j$ merge in the $(N+1)$-soliton solution, 
the expression eq.~(\ref{eq:Nsoliton}) 
for the solution remains true with replacing formally $\Pi_i\to\Pi_i+\Pi_j$. 
Recall that $\det(\Pi_i+\Pi_j)\neq0$ in general, even when 
$\det\Pi_i=0$ and $\det\Pi_j=0$. 

Finally, multiplying the carrier wave part $\e^{\im(kx-\omega t)}$ 
as eq.~(\ref{eq:changeQ}), 
the multi-soliton solution for the original NLS equation (\ref{eq:genNLS}) 
under the nonvanishing boundary conditions 
(\ref{eq:constBC}), (\ref{eq:constBC+}) is obtained: 
\begin{align}
Q(x,t)=\lambda_0\e^{\smalli(kx-\omega t)}
\left\{I+2\im(\overbrace{I\cdots I}^{N})S^{-1}\left[
\begin{array}{c}
\Pi_1\e^{\chi_1}\\
\vdots\\
\Pi_{N}\e^{\chi_{N}}
\end{array}
\right]\right\}.
\label{eq:finalNsoliton}
\end{align}
As mentioned in Sec.~\ref{ssec:NBC}, the multi-soliton solutions 
for general nonvanishing 
boundary conditions are easily obtained through the unitary transformations 
(\ref{eq:unitarytransf}).

\section{Concluding Remarks}
\label{sec:conclusions}
In this paper, we have studied both the self-defocusing and the self-focusing matrix nonlinear Schr\"odinger (NLS) equations under nonvanishing boundary conditions. Introducing the Lax pair, we have made the inverse scattering method (ISM) analysis for the systems and shown that the initial value problem is solvable. From the Gel'fand--Levitan--Marchenko equation with the reflection-free condition, the multi-soliton solution is obtained explicitly. The conservation laws are given in the same manner as for the vanishing boundary conditions~\cite{Tsuchida1}, leading to an infinite number of the conserved quantities. On the other hand, the Lax pair, the direct and inverse problems, and the multi-soliton solution are altered due to the boundary conditions.

The multicomponent systems with nonvanishing boundary conditions include several important models in physics, for example the spinor model for Bose--Einstein condensates with repulsive and antiferromagnetic interactions. The equation for the dynamics of $F=1$ spinor condensate falls into the $l=2$ case of the matrix NLS equation. As expected from the applicability of the matrix NLS equation to the spinor bright solitons~\cite{IMWlett,IMWfull}, it is interesting to analyze the spinor ``dark" solitons based on the results obtained in this work. We report this issue in a separate paper~\cite{UIW}.

\end{document}